\documentclass[conference]{IEEEtran}
\usepackage{graphicx,url}
\usepackage[caption=false,font=footnotesize, hangindent=10pt]{subfig}
\usepackage{amsmath,amsfonts,bm}
\usepackage{color}
\usepackage{multirow,tabularx,tabu}

\makeatletter
\long\def\@makecaption#1#2{\ifx\@captype\@IEEEtablestring%
\footnotesize\begin{center}{\normalfont\footnotesize #1}\\
{\normalfont\footnotesize\scshape #2}\end{center}%
\@IEEEtablecaptionsepspace
\else
\@IEEEfigurecaptionsepspace
\setbox\@tempboxa\hbox{\normalfont\footnotesize {#1.}~~ #2}%
\ifdim \wd\@tempboxa >\hsize%
\setbox\@tempboxa\hbox{\normalfont\footnotesize {#1.}~~ }%
\parbox[t]{\hsize}{\normalfont\footnotesize \noindent\unhbox\@tempboxa#2}%
\else
\hbox to\hsize{\normalfont\footnotesize\hfil\box\@tempboxa\hfil}\fi\fi}
\makeatother

\title{A Peep on the Interplays between Online Video Websites and Online Social 
Networks}

\author{
\IEEEauthorblockN{Junzhou Zhao, Pinghui Wang, Jing Tao, Xiaobo Ma and Xiaohong Guan}
\IEEEauthorblockA{MOE KLINNS Lab, Xi'an Jiaotong University, Xi'an, P.R. China}
\IEEEauthorblockA{Email: \{jzzhao, phwang, jtao, xbma, xhguan\}@sei.xjtu.edu.cn}
}

\begin{document}

\maketitle

\begin{abstract}
Many online video websites provide the shortcut links to facilitate the video 
sharing to other websites especially to the online social networks (OSNs). Such 
video sharing behavior greatly changes the interplays between the two types of 
websites. For example, users in OSNs may watch and re-share videos shared by their 
friends from online video websites, and this can also boost the popularity of 
videos in online video websites and attract more people to watch and share them. 
Characterizing these interplays can provide great insights for understanding the 
relationships among online video websites, OSNs, ISPs and so on. 

In this paper we conduct empirical experiments to study the interplays between 
video sharing websites and OSNs using three totally different data sources: online 
video  websites, OSNs, and campus network traffic. We find that, a) there are many 
factors that can affect the external sharing probability of videos in online video 
websites. b) The popularity of a video itself in online video websites can
greatly impact on its popularity in OSNs. Videos in Renren, Qzone (the top two most 
popular Chinese OSNs) usually attract more viewers than in Sina and Tencent Weibo 
(the top two most popular Chinese microblogs), which indicates the different
natures of the two kinds of OSNs. c) The analysis based on real traffic data 
illustrates that 10\% of video flows are related to OSNs, and they account for 25\% 
of traffic generated by all videos. 

\end{abstract}

\section{Introduction}

Video traffic is rapidly growing in the Internet. It is reported that 15\% to
25\% of all the inter-autonomous system traffic today is video\cite{Labovitz2010}. 
Recently, according to comScore's report released in February 
2013\cite{comScore2013}, besides the online video websites, online social networks 
(ONSs) such as Facebook are the second largest platforms for people to watch 
videos. Since the majority of these videos are still hosted in online video 
websites such as YouTube and Hulu. Hence, it is an interesting topic to study 
\emph{how these videos transfer to OSNs}, which also forms one of our motivations  
in this paper. 
 
\begin{figure}
\centering
\includegraphics[width=\linewidth]{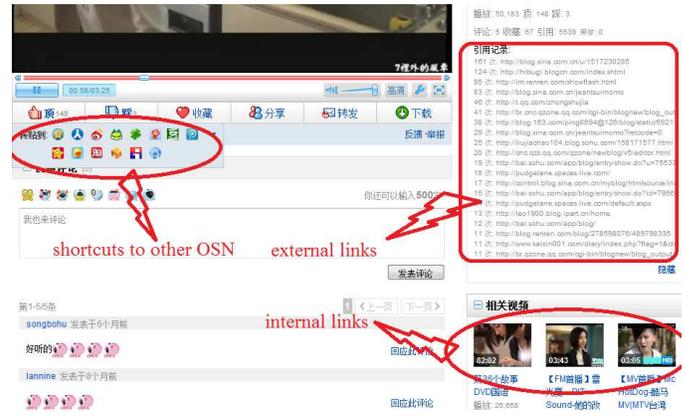}
\caption{A video profile in Youku.}
\label{fig:youku}
\end{figure}

In fact, many online video websites provide people the sharing buttons to 
facilitate the video sharing to other external websites such as Facebook, Twitter 
and personal blogs. An example is shown in Fig.~\ref{fig:youku}. Youku, a popular 
online video website in China, which is just as famous as YouTube in U.S., provides 
such video sharing function for each of its video. When a Youku user finds that 
a video is interesting, he can click on the sharing button, choose an OSN, then the 
video will be shared to that OSN immediately. At the same time, Youku provides 
brief summary information of external links that have referred this video. 

Besides Youku, there are many other online video websites in China competing
with each other, e.g., Tudou, Ku6.com, 56.com and so on. Many of them announced
that they are the top bananas of the online video industry of China, which
usually makes a third person confused to these unbelievable declarations.
Similar to online video websites, there are also many OSNs co-existing in
China, e.g., Renren, Douban, Qzone, Sina Weibo, Tencent Weibo and so on. Their
situations are just similar to or even worse than the online video websites.
Hence, there are requirements to \emph{bring orders to this disordered online
ecosystem}, which forms our second motivation to study them and provide people
a more realistic knowledge about their positions in the online ecosystem. 

In order to study the complex relationships between them, we conduct an in depth 
analysis based on datasets collected from multiple perspective of views of data 
sources: 
\begin{enumerate}
\item Data form online video websites, Youku and Tudou. 
\item Data from OSNs, Renren and Sina Weibo.
\item Network traffic data from a campus network. 
\end{enumerate}

We summarize our findings as follows:
\begin{itemize}
\item Based on the dataset collected from online video websites, we find that
many factors can affect the external sharing probability of videos in online
video websites. The nature of a video, e.g., its category, is also an important
factor. 

\item Based on the dataset collected from Renren and Sina Weibo, we find that
the popularity of a video itself in online video websites can greatly impact on
its popularity in OSNs. Videos in Renren and Qzone usually attract more viewers
than in Sina and Tencent Weibo, which indicates the different natures of the
two kinds of OSNs.

\item The analysis based on real traffic data shows that 10\% of video flows
are associated with OSNs, and they account for 25\% of traffic generated by all
videos.
\end{itemize}

The outline of this paper is as follows. Section ~\ref{sec:related} summarizes
related works. The data collection process is described in Section
~\ref{sec:data}. Section ~\ref{sec:analysis} presents the results of an
in-depth analysis of the collected datasets. Concluding remarks then follow.

\section{Related Work}\label{sec:related}
Cha et al.~\cite{Cha2007} crawled the YouTube and Daum UCC, the most popular UGC 
service in Korea, and presented an extensive analysis of the video popularity 
distribution, popularity evolution, user behavior analysis, and content duplication 
in YouTube. Further, Cheng et al.~\cite{Cheng2008} investigated the social networks 
in YouTube videos. Chatzopoulou et al.~\cite{Chatzopoulou2010} analyzed popularity 
by looking at properties and patterns in time and considering various popularity 
metrics, and studied the relationship of the popularity metrics. \cite{Gill2007} 
and \cite{Zink2009} collected traces at the edge of a single campus network and 
studied usage patterns, video properties, popularity and referencing 
characteristics, and transfer behaviors of YouTube from the perspective of an edge 
network. Saxena et al.~\cite{Saxena2008} analyzed and compared the underlying 
distribution frameworks of three video sharing services---YouTube, Dailymotion and 
Metacafe, based on traces collected from measurements performed in a PlanetLab 
environment.

Using traffic trace collected at multiple PoPs of the ISP, Adhikari et 
al.~\cite{Adhikari2010} inferred the load balancing strategy used by YouTube to 
serve user requests. Torres et al.~\cite{Torres2011} employed state-of-the-art 
delay based geolocation techniques to find the geographical location of YouTube 
servers, and performed analysis on groups of related YouTube flows. Compared with a 
location-agnostic algorithm to map users to data centers analyzed in 
\cite{Adhikari2010}, they found that the YouTube infrastructure has been completely 
redesigned and now primarily uses a nearest RTT mapping policy. 

In \cite{Lai2009,Lai2010}, the authors studied the impacts of external links on the 
YouTube and Youku videos. Our work differs itself by focusing on relationships 
between online video websites and OSNs. And we use data from three totally 
different sources to support this study. 

\section{Data Collection Methods}\label{sec:data}

In this section, we describe the data collection methods from three different data 
sources. 

\subsection{Data Collection from Youku and Tudou}

Youku\footnote{http://www.youku.com} and Tudou\footnote{http://www.tudou.com. Tudou
was acquired by Youku in March 2012.} are the top two most popular online video
websites in China. To collect videos in Youku, previous study\cite{Lai2010} uses 
the simple Breadth-First-Search (BFS) method. However it is known that incomplete 
BFS is likely to densely cover only some specific region of the Youku video 
network, which can introduce uncorrectable statistic 
bias\cite{Achlioptas2005,Kurant2010}. Hence, results obtained from data collected 
by BFS is unbelievable. Fortunately, we find that Youku assigns an eight digits 
numeric ID to each videos, so a uniform sample of Youku videos can be obtained by 
generating uniformly random numbers, and by polling Youku to known about their 
existences. Based on this uniform video sampling method, 1.5 millions of videos in 
Youku are collected. Similar data collection process is also conducted in Tudou, in 
which we also uniformly collected 1.5 millions videos. For each video in Youku and 
Tudou, its profile information is also retrieved, which is shown in 
Table~\ref{tab:video_data}.

\begin{table}
\centering
\caption{Video profiles in Youku and Tudou. A tick represents having this item, and 
a cross represents not having this item.}
\label{tab:video_data}
\begin{tabular}{l|c|c}
\hline\hline
Profile & Youku & Tudou \\
\hline
category			& $\surd$ 	& $\surd$ \\
length (minutes)	& $\surd$ 	& $\surd$ \\
size (bytes)		& $\surd$ 	& $\times$ \\
rating				& $\times$	& $\surd$ \\
uploaded date 		& $\surd$ 	& $\surd$ \\
\#views				& $\surd$ 	& $\surd$ \\
\#comments 		& $\surd$ 	& $\surd$ \\
\#favorites 		& $\surd$ 	& $\surd$ \\
\#likes   		& $\surd$ 	& $\surd$ \\
\#dislikes 		& $\surd$ 	& $\surd$ \\
\#external shares 	& $\surd$ 	& $\surd$ \\
top 20 external shared links 	& $\surd$ & $\surd$ \\
\hline
\end{tabular}
\end{table}

\subsection{Data Collection from Renren and Sina Weibo}

Renren\footnote{http://www.renren.com} is one of the largest OSNs in China with 
more than 150 million users. Every user can watch or share videos in Renren. This 
sharing can be external (from online video websites to Renren) or internal 
(inside Renren from a user to another user). We collect all the Renren accounts 
related to Xi'an Jiaotong University (they or their friends are in the campus), 
which form a network of 1,661,236 nodes and 32,050,611 edges. From these users, 
we further collect 0.5M videos. For each video, we record its original URL link 
address, the number of views and shares inside Renren. 

We also collect data from Sina Weibo\footnote{http://www.weibo.com}, one of the 
largest microblogs in China, which has more than 200 million users. Sina Weibo 
assigns a ten digits numeric ID to each user, therefore we can collect tweets from 
14,623 uniform randomly sampled users and about 3.7M tweets are collected. Each 
tweet is classified to a retweet (posted by retweeting an existing tweet) or an 
original tweet (posted by self writing). About 64.8\% of the tweets are retweets in 
our dataset. 

\subsection{Traffic Collection from the Edge}
The traffic data used in this paper is based on the actual network traffic over
the backbones of CERNET (China Education and Research Network) Northwest
Regional Center and the campus network of Xi'an Jiaotong University. The
traffic data is collected at an egress router with a bandwidth of 1.5 Gbps by
using TCPDUMP for about ten days in March 2011. Since application level
characteristics are our primary interests in this paper, our analysis focuses on
the HTTP traffic data. We group packets into different bidirectional
HTTP flows, where a flow is defined as a bidirectional, ordered
sequence of packets generated by a pair of (packet source IP, packet source
port) and (packet destination IP, packet destination port). For each HTTP
bidirectional flow, we record its starting time, finishing time, total bytes,
total packets, HTTP request (e.g, the method, URL, Host), and HTTP response
(e.g. status code, content type, content size). 40M HTTP flows are
collected, which include 13K distinct sources\footnote{All recorded
IP addresses are anonymized to protect users' privacy.}.

\section{Data Analysis}\label{sec:analysis}

In this section, we analyze the relationships between online video websits and 
OSNs from three different perspective of views. 

\subsection{External Video Sharing in Youku and Tudou}

\begin{figure}
	\centering
	\includegraphics[width=.6\linewidth]{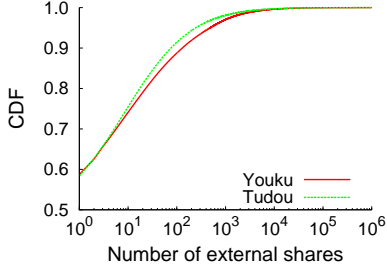}
	\caption{Cumulative distribution of the number of external shares.}
	\label{fig:yt_shares}
\end{figure}

First, we provide overview statistics of the video external shares in Youku and 
Tudou. Fig.~\ref{fig:yt_shares} depicts the empirical cumulative distributions 
(CDF) of the number of external shares for videos in Youku and Tudou. We find that 
the majority of videos are not widely shared externally. For example, 80\% of Youku 
videos have less than 22 shares, and 80\% of Tudou videos have less than 
18 shares. About 0.3\% of the videos in Youku and Tudou have more than 10,000 
external shares, and these minority videos are widely shared outside of Youku and 
Tudou. 

\begin{figure}
\centering
\subfloat{\label{fig:views}\includegraphics[width=.33\linewidth]{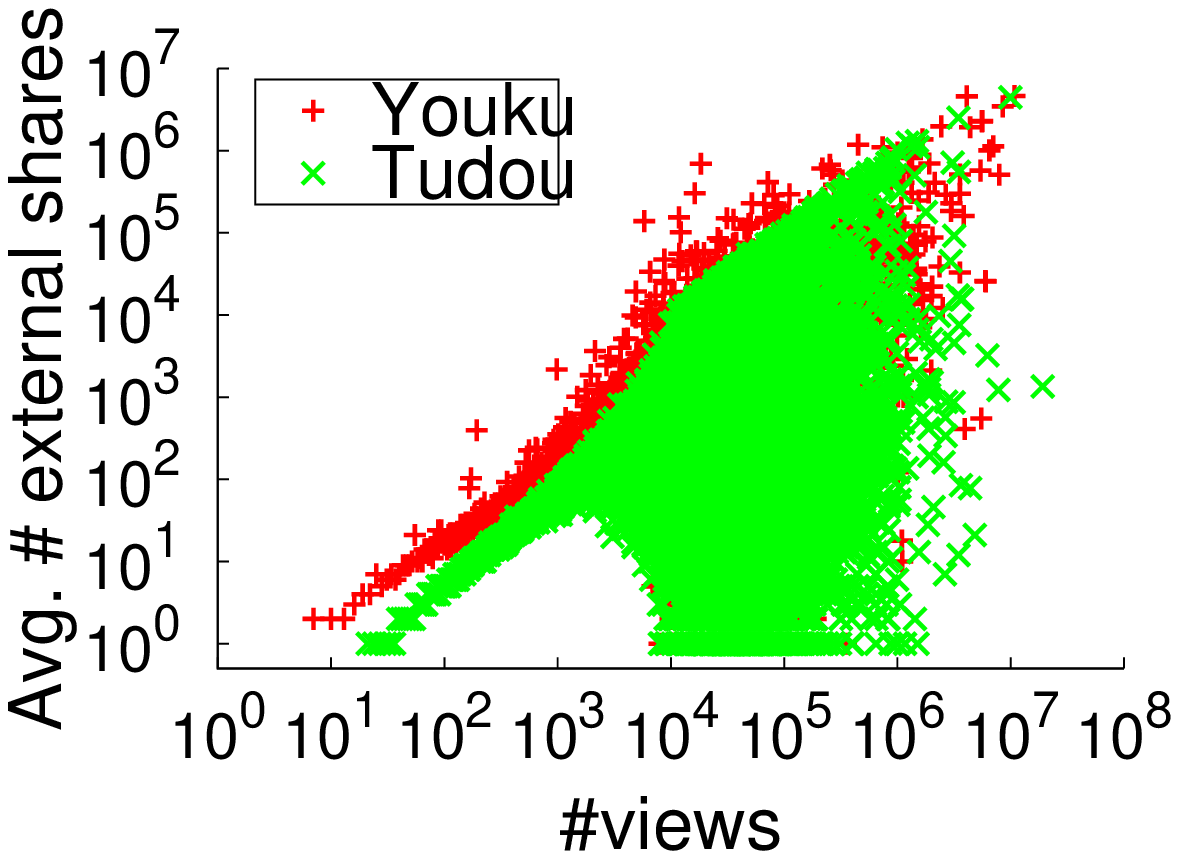}}
\subfloat{\label{fig:commt}\includegraphics[width=.33\linewidth]{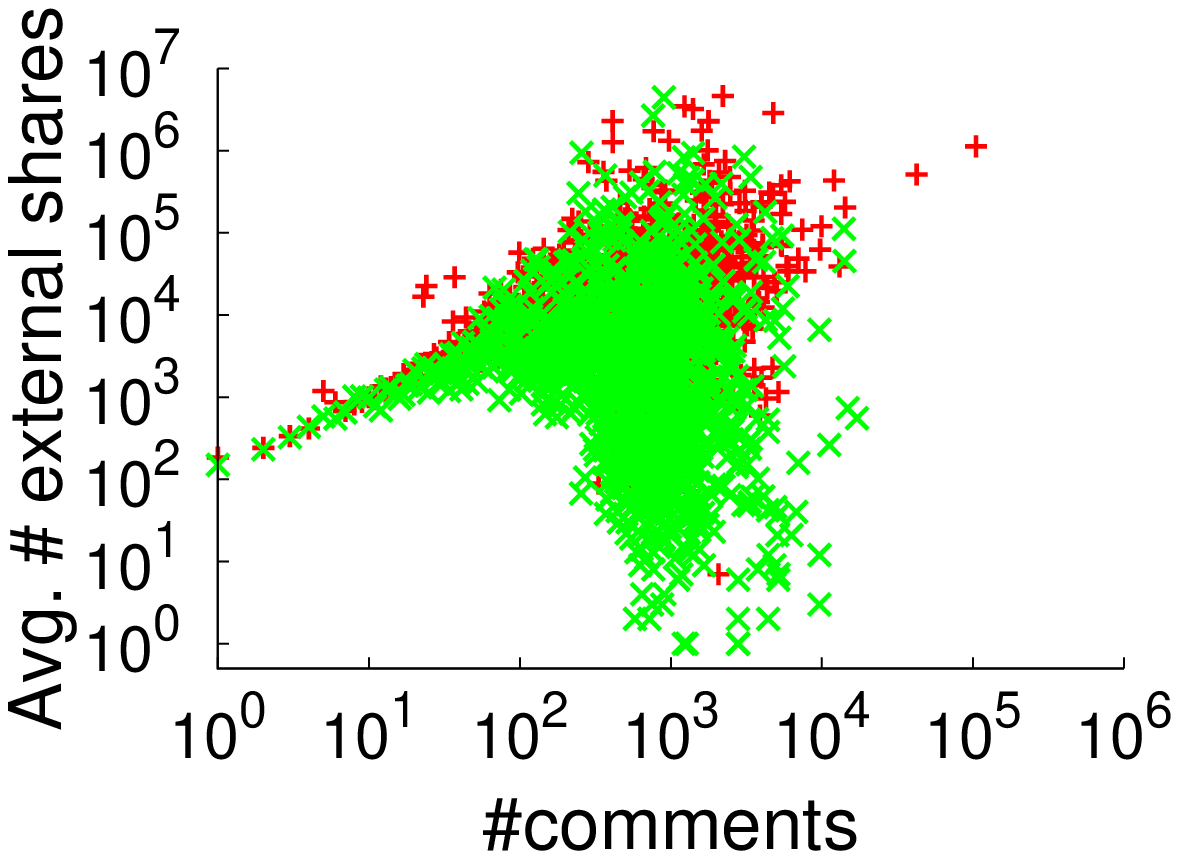}}
\subfloat{\label{fig:favat}\includegraphics[width=.33\linewidth]{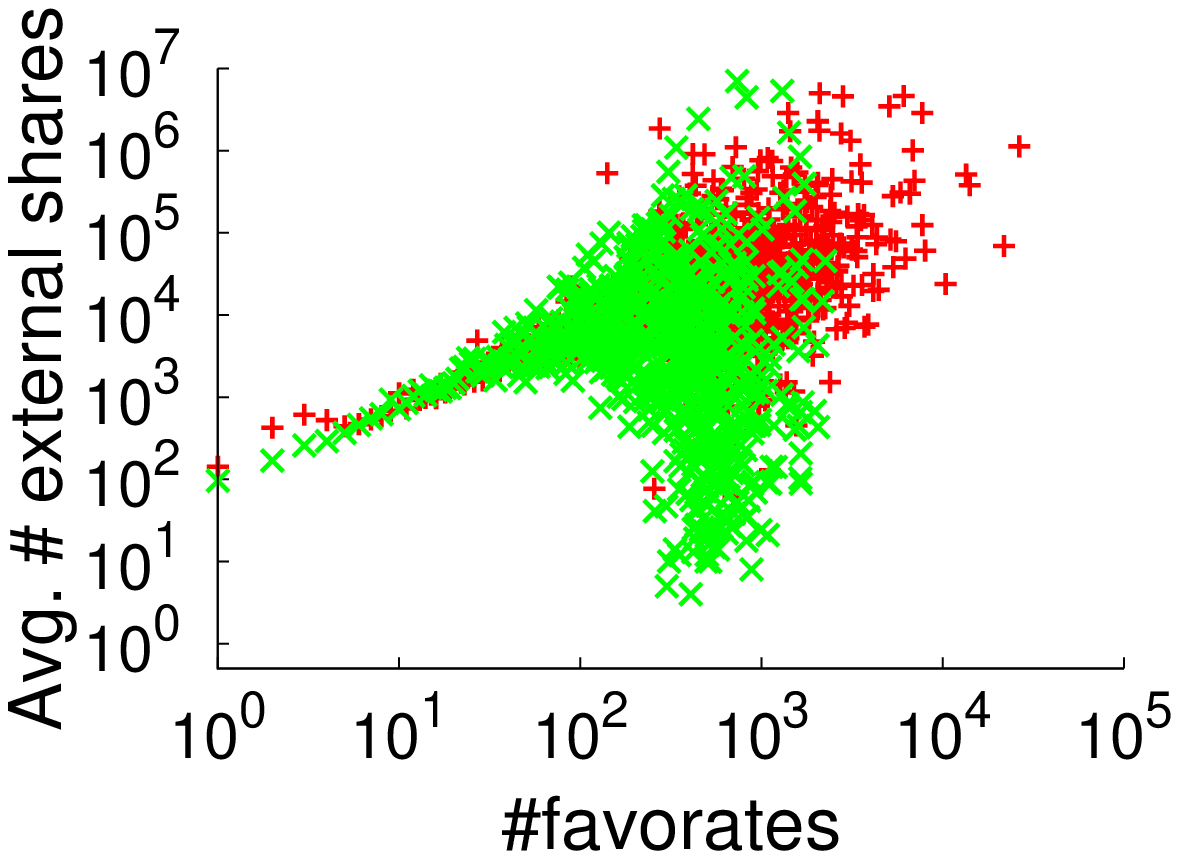}}

\subfloat{\label{fig:likes}\includegraphics[width=.33\linewidth]{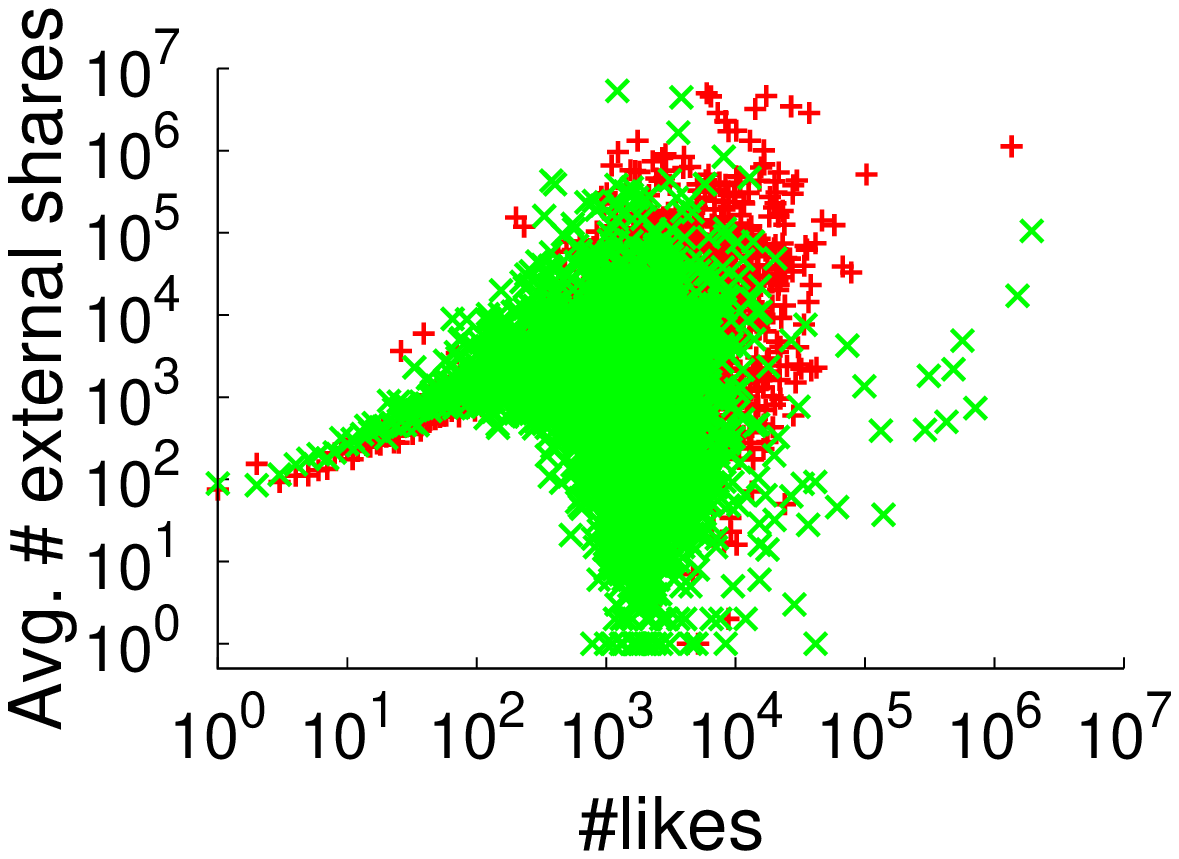}}
\subfloat{\label{fig:dislk}\includegraphics[width=.33\linewidth]{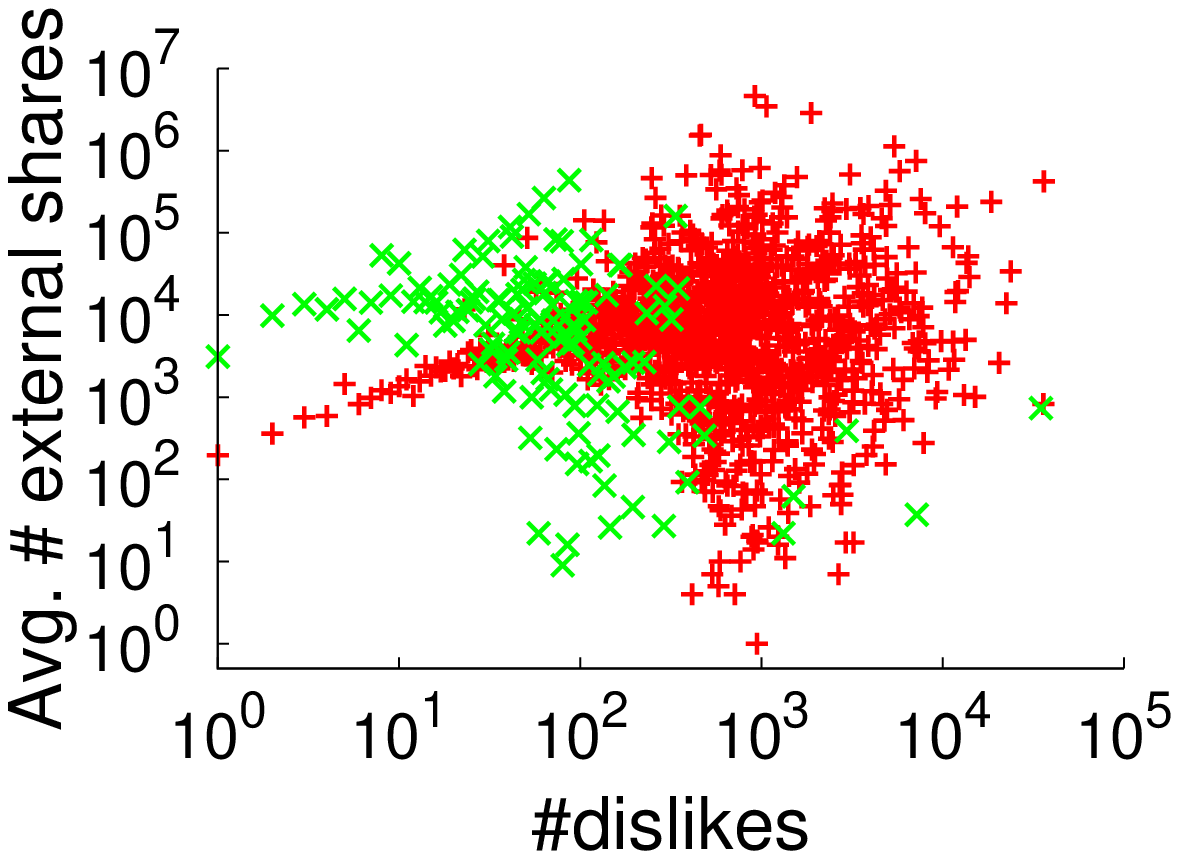}}
\subfloat{\label{fig:lengt}\includegraphics[width=.33\linewidth]{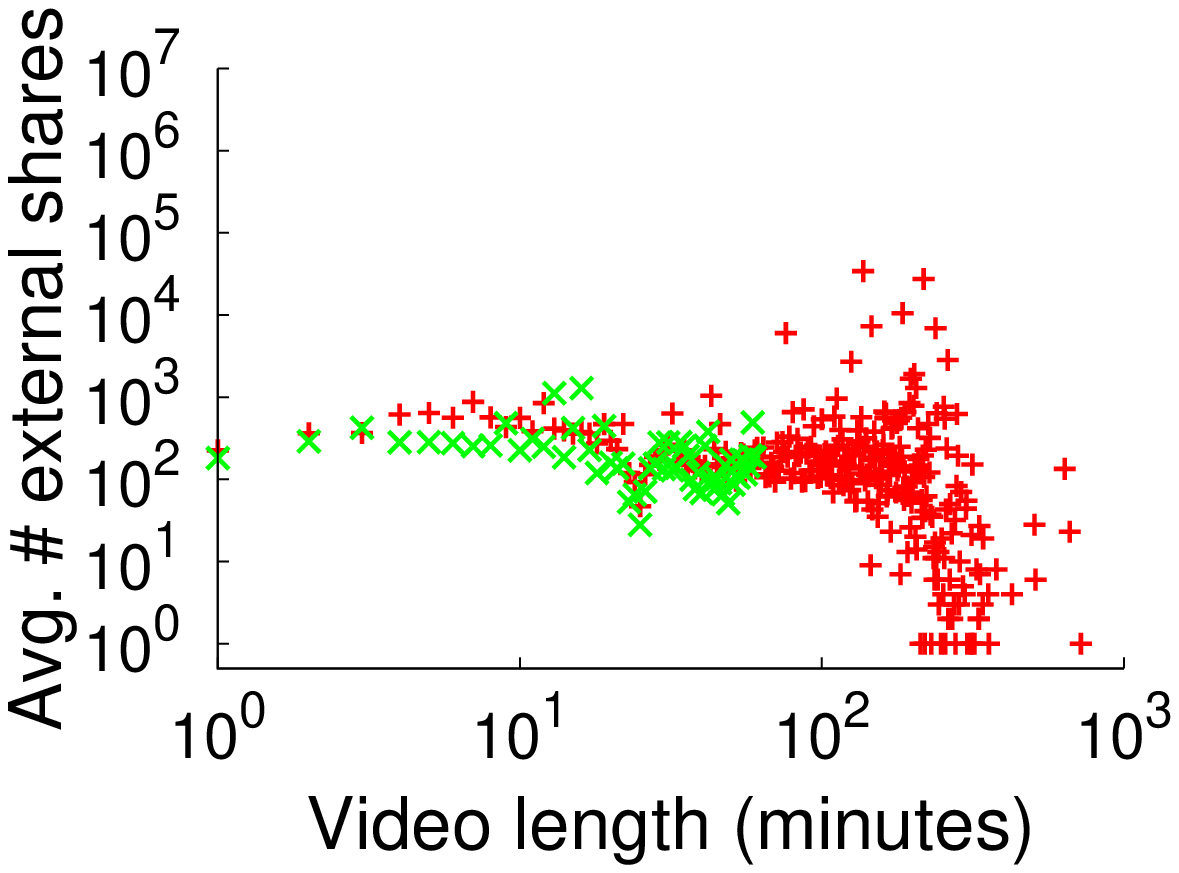}}

\subfloat{\label{fig:age}\includegraphics[width=.33\linewidth]{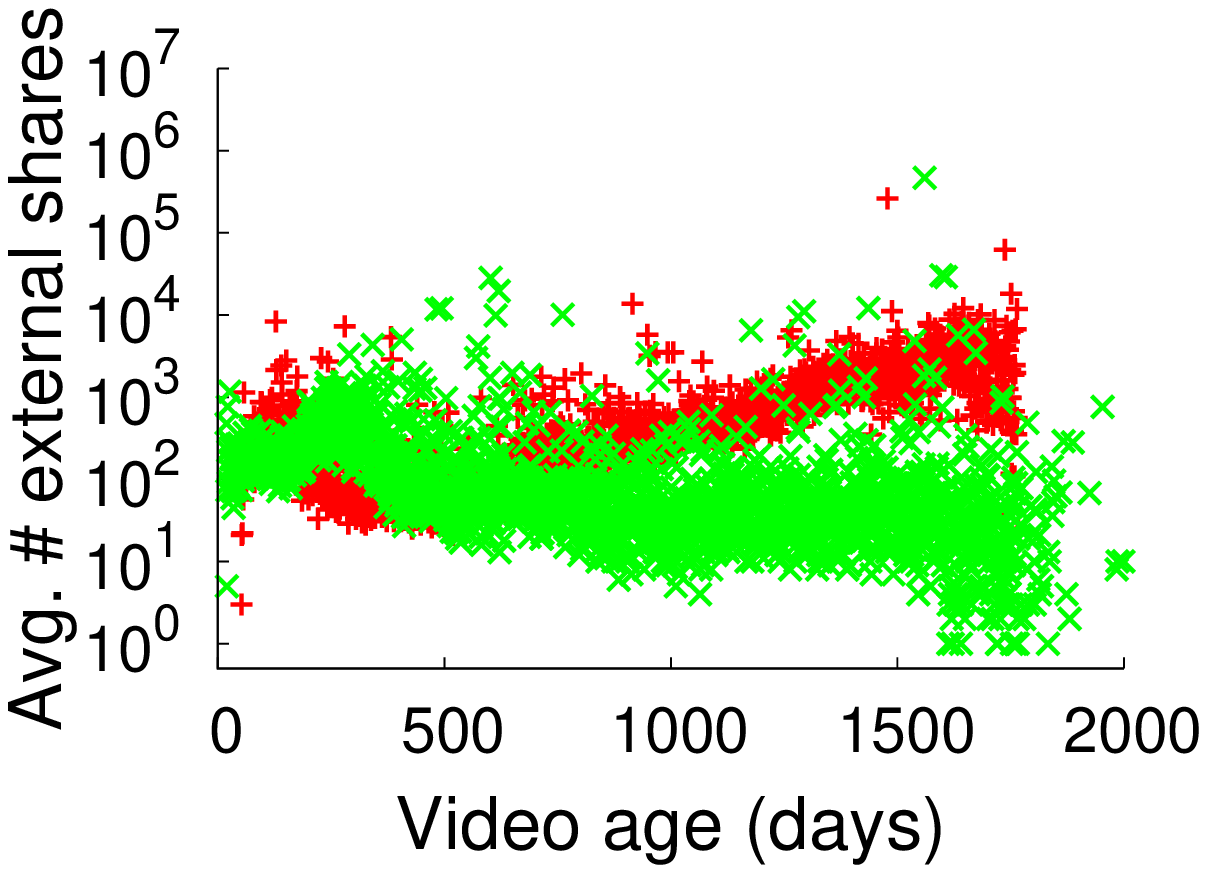}}
\subfloat{\label{fig:bit}\includegraphics[width=.33\linewidth]{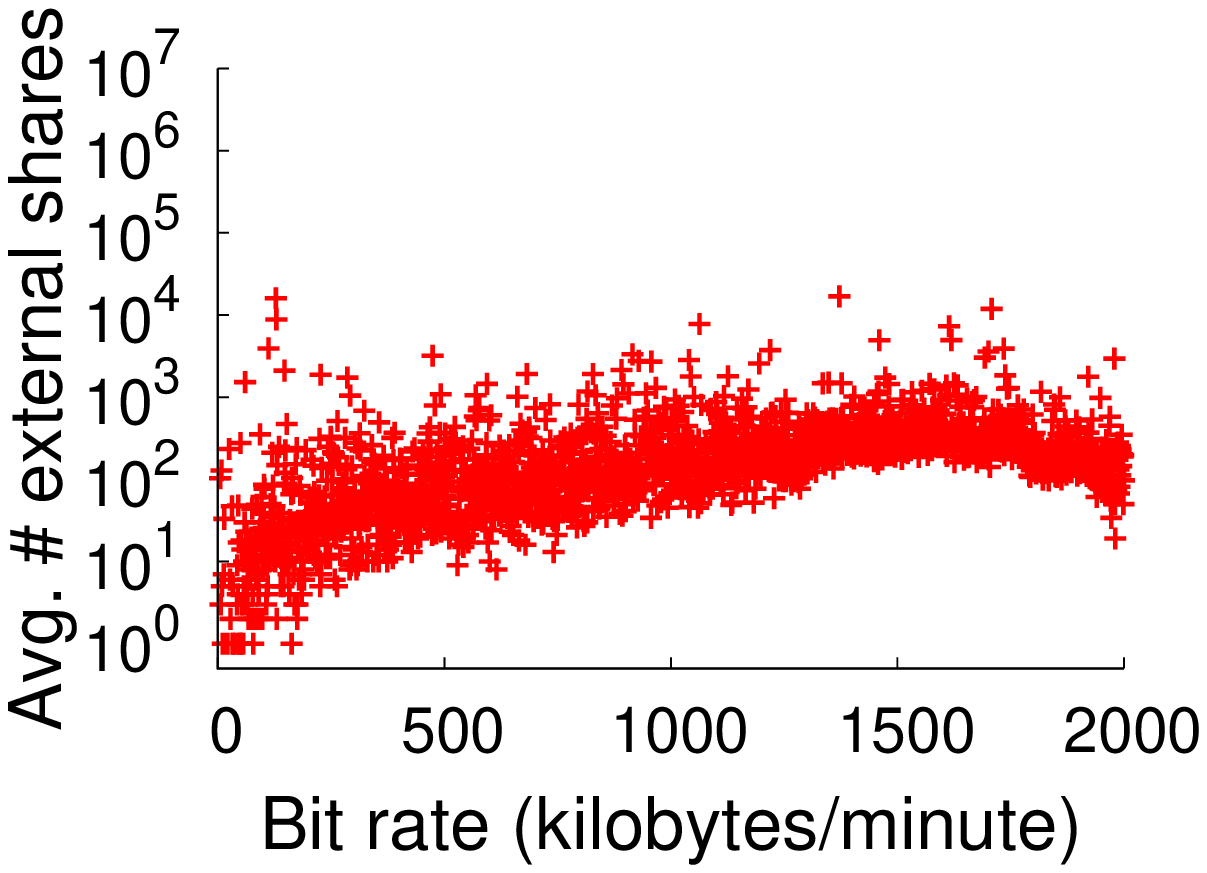}}
\subfloat{\label{fig:rat}\includegraphics[width=.33\linewidth]{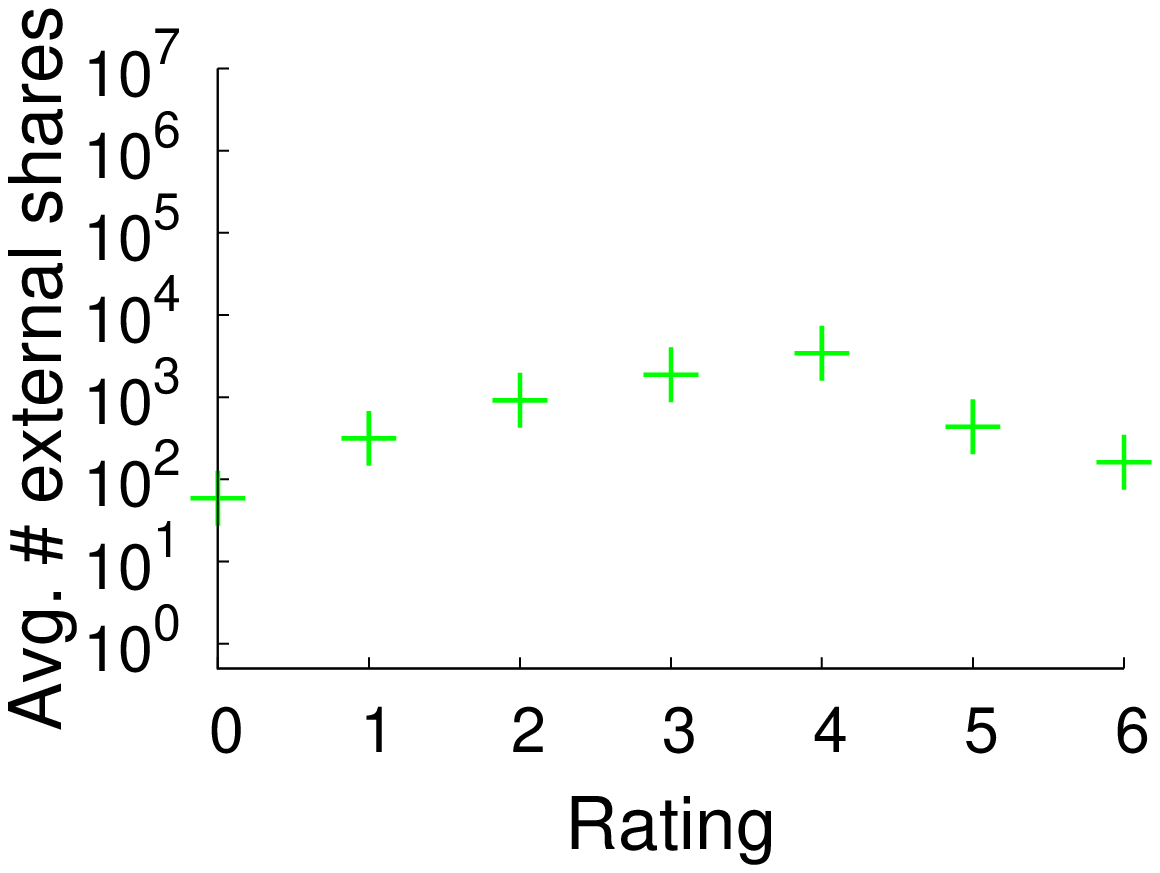}}
\caption{Relationship between video profiles and external shares.}
\label{fig:factors}
\end{figure}

Second, we want to study what are the factors that may cause a video to be widely 
shared outside of Youku and Tudou? Hence, we consider the video profiles 
listed in Tab.~\ref{tab:video_data}. The relationships between these factors 
and the average number of external shares are shown in Fig.~\ref{fig:factors}. Most 
of these factors are consistent with each other either in Youku or Tudou. We will
analyze them in detail. 

\noindent\textbullet \textbf{\#views} The number of views seems to have the 
strongest relationship with the possibility of external sharing of a video. 
Intuitively, a video receiving more views indicates itself is popular, hence it 
will be shared to external sites with a high probability. And because of the 
feedback effect, many external shares will also increase its views inside of a 
video site. 

\noindent\textbullet\textbf{\#comments} The number of comments has similar effect 
with the number of views, and it also has strong relationship with the external 
sharing. However, to comment on a video needs the user to login first, which may 
stop a lot of people who don't own accounts. 

\noindent\textbullet\textbf{\#favorites} When a video is interesting to a user, he 
can mark it as a favorite video (which results in that the video is saved in 
his personal homepage in the video site to facilitate his future watching). Since 
marking a video as favorite also needs the login, which may also reduce the 
correlation with external sharing. 

\noindent\textbullet\textbf{\#likes and \#dislikes} Comparing the two factors, a 
video receiving more likes will be more likely to be shared than a video has less 
likes or many dislikes. 

\noindent\textbullet\textbf{Length} Video length has weak correlation with video's 
external sharing. Comparing short videos with long videos, short videos have 
higher probability to be shared. This is because users may not have enough time to 
watch a long video and decide to share it. 

\noindent\textbullet\textbf{Age} Age also has weak correlation with a video's 
external sharing. Videos in Tudou shows that old videos are less likely to be 
shared than new videos. However, this is a little different in Youku. 

\noindent\textbullet\textbf{Bit rate and rating} Bit rate of a video is defined 
as Size$/$Length. Both the two factors can be used to characterize the quality of a 
video. Generally, higher quality videos has higher probability to be shared. 
However, this relationship is very weak. 

We use Pearson correlation coefficients (PCC) to quantify these correlations in 
Youku and Tudou, which are shown in Table \ref{tab:pearson}. PCC is the most common 
metric to measure the dependence between two quantities. For a pair of metrics $x$ 
and $y$, its corresponding values for the sampled video $i$ ($i=1,2,\ldots,n$) are 
$x_i$ and $y_i$, where $n$ is the total number of sampled videos. Then their PCC is 
defined as
\[
\rho_{xy}=\frac{\sum_{i=1}^{n}(x_i-\overline{x})(y_i-\overline{y})}
{\sqrt{\sum_{i=1}^{n}(x_i-\overline{x})^2\sum(y_i-\overline{y})^2}}
\]
where $\overline{x}=\sum_{i=1}^{n}x_i$ and $\overline{y}=\sum_{i=1}^{n}y_i$ are the 
sample means of $x$ and $y$. The results of PCCs are consistent with our analysis 
above. 

\begin{table}[htp]
	\centering
	\caption{PCCs of different factors.
	\label{tab:pearson}}
	\begin{tabular}{l|rr}
		\hline
		\hline
		Factors & Youku & Tudou \\
		\hline
		\#views	& 0.50	& 0.54 \\
		\#comments & 0.17 & 0.16 \\
		\#favorites & 0.14 & 0.15 \\
		\#likes  & 0.06 & 0.05 \\
		\#dislikes & 0.01 & 0.04 \\
		length & 0.00 & 0.00 \\
		age & 0.01 & 0.00 \\
		bit rate & 0.00 & - \\
		rating & - & 0.03 \\
		\hline
	\end{tabular}
\end{table}

\begin{figure}
\centering
\includegraphics*[width=.8\linewidth]{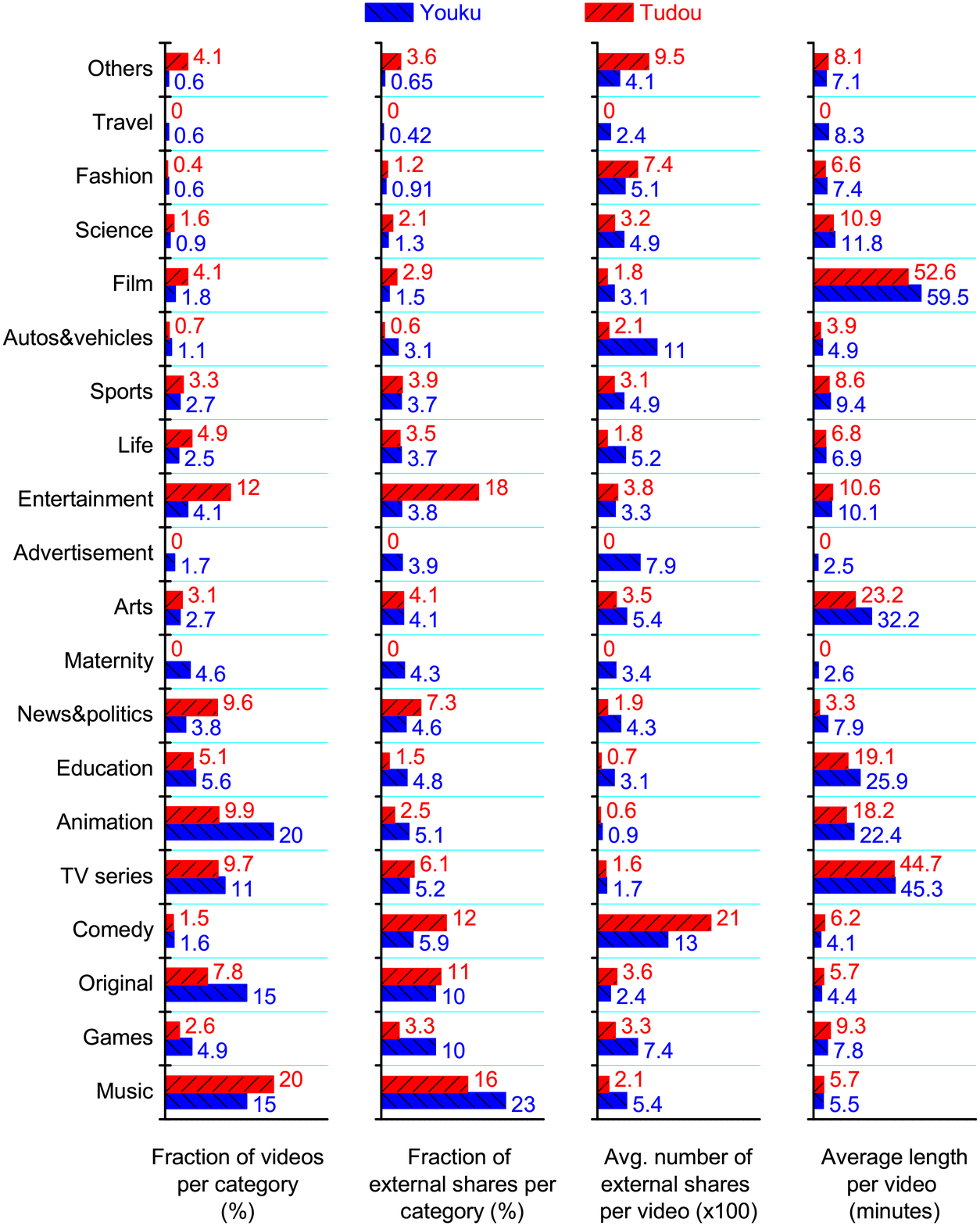}
\caption{Categories of videos in Youku and Tudou. \label{fig:cat}}
\end{figure}

Next, the category of a video also affects the external sharing a lot. To see 
this, we show category's effect in Fig.~\ref{fig:cat}. We find that music and 
animation are the top-2 categories that have the highest percentages of videos for 
both Youku and Tudou. Most external links in Youku come from video categories 
music, games, original, comedy, TV series, and the same top-5 categories in Tudou 
come from categories entertainment, music, comedy, original and Hot topic. The 
largest averages of the number of external shares per video are both comedy for 
Youku and Tudou.

Finally, we study what are these external links pointing to, or what external  
sites are consuming these videos. Table ~\ref{tab:yk_t20} states the most 
frequently external websites identified from the external links in Youku and Tudou. 
Among the external websites, Renren and Qzone consume the majority of the videos, 
about 96\% of the Youku videos and 75\% of the Tudou videos. Meanwhile, Renren and 
Qzone consume more videos than microblogs such as Sina and Tencent Weibo. 

\begin{table}
\centering
\caption{Top external OSNs consuming Youku and Tudou videos.}
\label{tab:yk_t20}
\begin{tabular}{l|rr}
	\hline
	\hline
	\multirow{2}{*}{External site} & \multicolumn{2}{c}{\#external links 
	($\times10^5$)} \\ 
	\cline{2-3}
	& Youku & Tudou \\
	\hline
	Renren.com	& 110 & 220\\
	Qzone.qq.com& 470 & 120\\
	Kaixin001.com	& 12 & 99\\
	Douban.com	& 0.27 & 1.1\\
	Sina Weibo (weibo.com) & 5.0 & 4.7\\
	Tencent Weibo (t.qq.com) & 6.4 & 8.1\\
	\hline
\end{tabular}
\end{table}

\subsection{Video Consuming in Renren and Sina Weibo}

Having understand the external video sharing in online video websites, we now 
move to study how videos are consumed in Renren and Sina Weibo, which are the 
two most popular ONSs in China (as famous as Facebook and Twitter in U.S.). When a 
video is 
shared from an online video website to an OSN, the remaining behaviors related to 
this video is watching and sharing within the OSN. Hence, in the following, if we 
don't clarify, the sharing behavior means internal sharing within an OSN, which is 
different from the external sharing studied in previous. 

\begin{figure}
	\centering
	\subfloat[CDF of \#shares/\#views.\label{fig:rr_shares}]
	{\includegraphics[width=.5\linewidth]{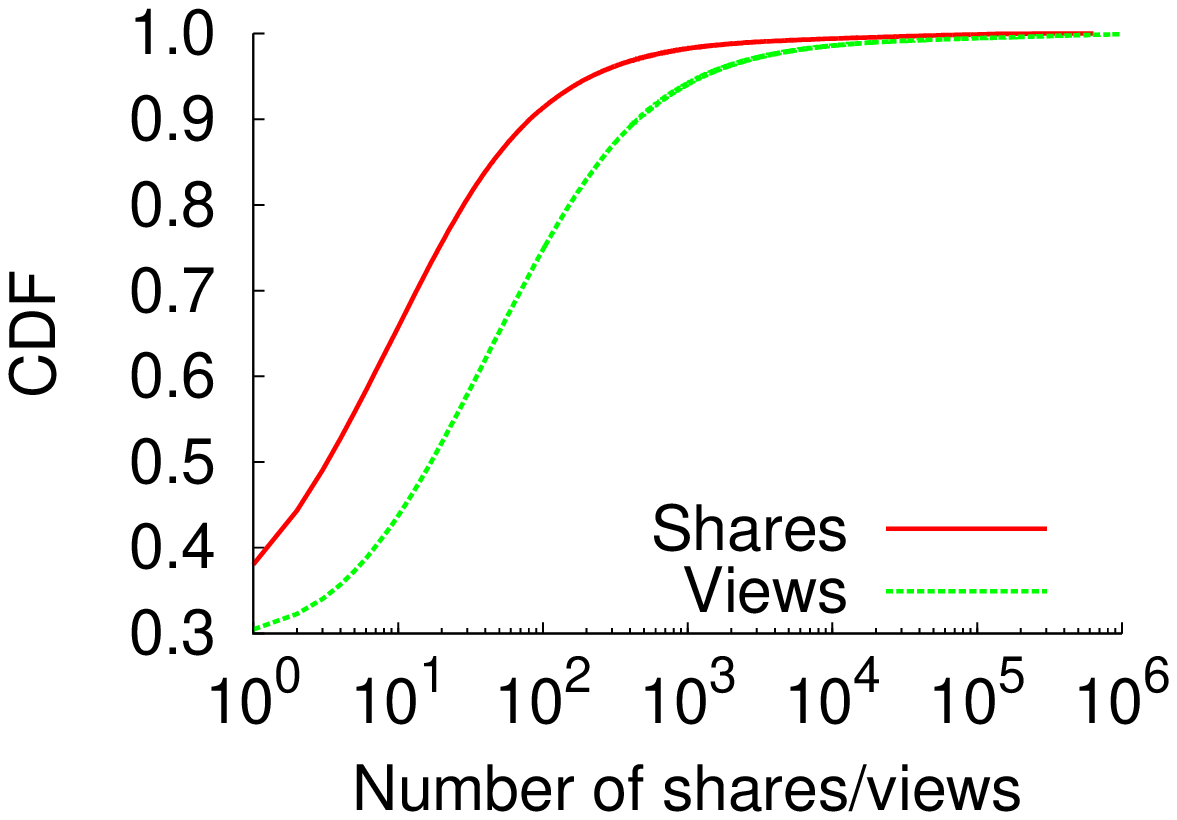}}
	\subfloat[Relationship between \#views and \#shares.\label{fig:rr_relation}]
	{\includegraphics[width=.5\linewidth]{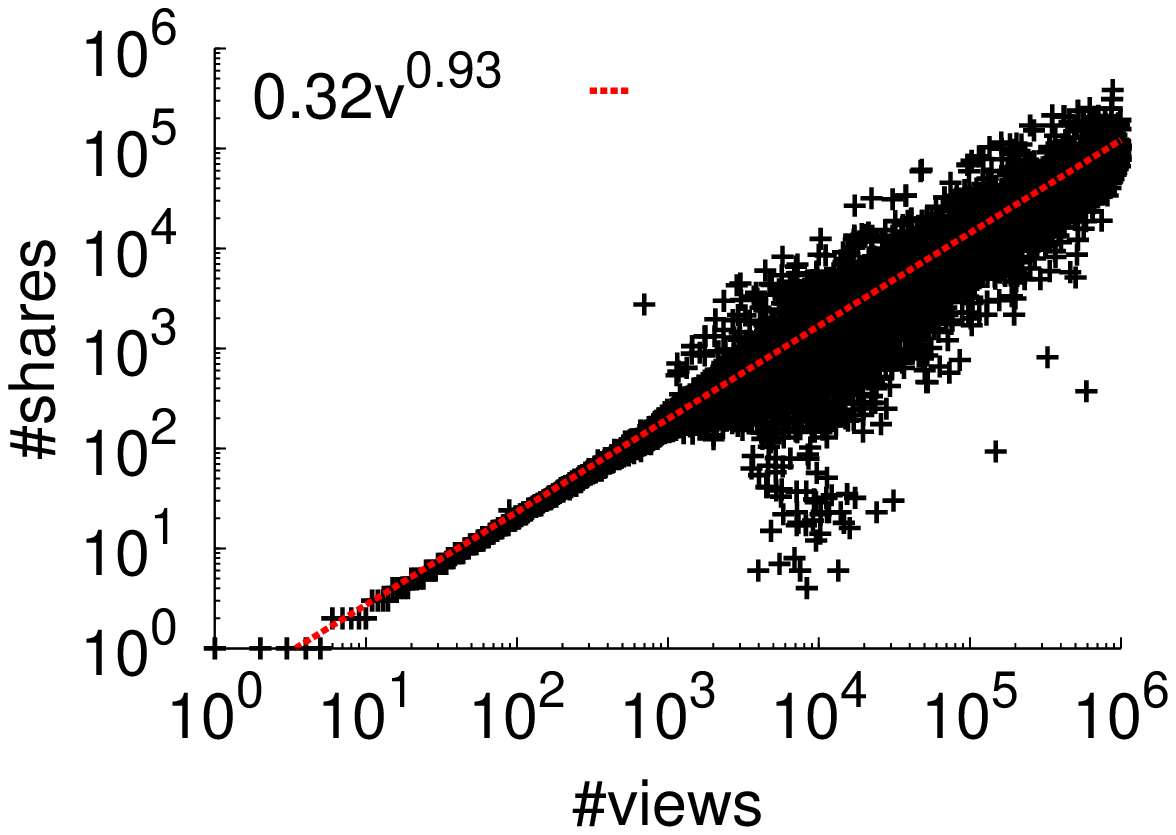}}
	\caption{Videos in Renren.\label{fig:rr_shares}}
\end{figure}

First, we provide a general picture about video views and shares in Renren. 
Fig.~\ref{fig:rr_shares} shows the CCDF of the number of shares and 
views for a video. We find that the majority of videos are not shared or watched by 
a lot of people, e.g., 80\% of videos are shared less than 29 times, and 80\% of 
videos are viewed by Renren users less than 153 times, which is similar to the 
external video sharing in online video websites. 
They both have a long heavy tail. When we further investigate the relationship 
between views and shares, we find it has a very simple mathematic formula as shown 
in Fig.~\ref{fig:rr_relation}. If denote the number of views by $v$, and the 
number of shares by $s$, then the relationship between $v$ and $s$ is 
\[
l=0.32v^{0.93},
\]
which is very close to a linear relationship. 

Next, we do a reverse engineering study to study where the videos come from. The 
answer to this question can give us a rank of the popularity of online video 
websites. We summarize the most popular video websites in Renren in 
Tab.~\ref{tab:rr_cat}. In fact, using different popularity metrics will obtain 
different ranks. We find that videos from Youku have the largest amount, views and 
shares, e.g., 57\% of the videos in Renren are from Youku. These videos attract 
71\% of all the views and 66\% of all the shares. However, when averaged to each 
video, videos from Ku6.com has the highest average views and shares, e.g., each 
video in Ku6.com can attract 1102 shares and 11215 views on average. Since videos 
from Ku6.com only account for 4.8\%, the average figures indicate that these 
minority videos are high quality. For videos from each video website, the average 
number of views per share has much smaller variance than the average number of 
views per video, which indicates that the number of views is strongly related to 
the number of share actions. 

\begin{table*}
\centering
\caption{Online video websites of videos shared in Renren.\label{tab:rr_cat}}
\begin{tabu}{l|X[r]X[r]X[r]X[r]X[r]X[r]}
	\hline
	\hline
	Website & 
	Fraction of views per website (\%) & 
	Fraction of shares per website (\%) & 
	Fraction of videos per website (\%) & 
	Avg. shares per video per website & 
	Avg. views per video per website & 
	Avg. views per share per website \\
	\hline
	Youku & 71 & 66 & 57 & 522 & 3942 & 7.5\\
	Ku6.com & 12 & 16 & 4.8 & 1102 & 11215 & 10\\
	Tudou & 9.2 & 9.6 & 26 & 150 & 1262 & 8.4\\
	56.com & 2.3 & 2.6 & 6.3 & 151 & 1404 & 9.3\\
	Sina.video.com & 1.4 & 1.8 & 0.95 & 618 & 6497 & 11\\
	Joy.cn & 1.0 & 1.5 & 0.93 & 463 & 5607 & 12\\
	Yinyuetai.com & 0.95 & 1.0 & 1.2 & 337 & 2932 & 8.7\\
	Tv.sohu.com & 0.40 & 0.73 & 0.88 & 191 & 2854 & 15\\
	Others & 1.0 & 0.81 & 1.9 & 225 & 1485 & 6.6\\
	\hline
\end{tabu}
\end{table*}

Now we move to Sina Weibo, the largest microblog in China. 
In Sina Weibo, about 7.4\% of the tweets contain a video and the same number
is 2.4\% in the original tweets, 11.2\% in the retweets. About 85.8\% of the video 
tweets are retweets. These figures illustrate that people are more reluctant to 
retweet a video tweet than uploading a video by himself. 

Weibo users can post tweets via various ways. Loginning into an account from a 
browser and posting tweets is the most common way. Except that, about 27.2\% of the 
users post tweets via mobile devices. And the top 10 most popular mobile OSs or 
devices used by Sina Weibo users are iPhone(13.92\%), Android(5.62\%), 
Nokia(4.71\%), iPad(1.72\%), Java(1.30\%), Motorola(0.19\%), HTC(0.15\%), 
BlackBerry(0.14\%), SonyEricsson(0.10\%), and WindowsMobile(0.08\%). There is a 
difference of posting behavior for video tweets and non-video tweets. For non-video 
tweets, the proportion of postings via mobile devices is 27.4\%. But for video 
tweets, the number is only 0.94\%. This indicates that mobile devices are still not 
convenient for posting video tweets. 

\begin{figure}
	\centering
	\includegraphics[width=.8\linewidth]{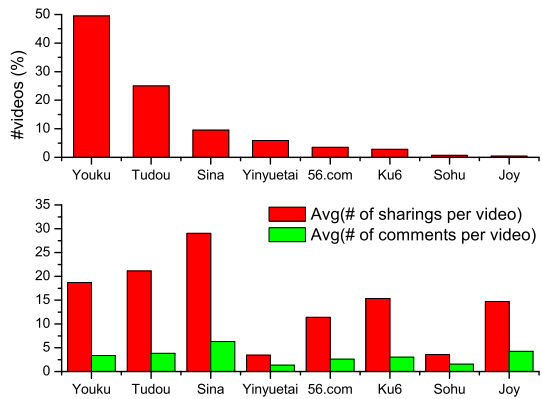}
	\caption{Video sites in Sina Weibo}
	\label{fig:video_weibo}
\end{figure}

Fig.~\ref{fig:video_weibo} shows the top 8 most popular video sites in Sina
Weibo. Youku and Tudou are the top 2 video sites which generate more than 
70\% of the videos in Sina Microblog. But the most retweeted and commented
videos are came from Sina itself, which takes account only 9.5\% of the videos.
This indicates that videos from Sina are more attractive than others. 

\subsection{A Peep from Campus HTTP Traffic Data}
Finally, we study the interactions between online video websites and OSNs from a 
much lower level---the campus HTTP traffic. The Internet traffic information should 
also reflect some properties of these interactions. To see this, we first show the 
fraction of different traffic from a campus in Table~\ref{tab:traff_frac} and 
Table~\ref{tab:video_traff}. We can see that 32\% of the traffic is made of video, 
and 39.5\% of the video traffic comes from Youku. The videos from Sohu have the 
highest download speed. 

\begin{table}
\centering
\caption{Traffic statistics by content type.\label{tab:traff_frac}}
\begin{tabu}{l|X[r]X[r]}
	\hline\hline
	& Distribution of \#bytes (\%) & Distribution of \#flows (\%)\\
	\hline
	Videos & 32.0 & 18.8 \\
	Images & 31.5 & 32.8 \\
	Text   & 11.2 & 9.5 \\
	Applications & 2.2 & 32.9 \\
	Others & 22.8 & 6.1 \\
	\hline
\end{tabu}
\end{table}

\begin{table}
\centering
\caption{Traffic statistics by online video websites.\label{tab:video_traff}}
\begin{tabu}{l|X[r]X[r]X[r]}
	\hline\hline
	Website & 
	Distribution of \#bytes (\%) &
	Distribution of \#flows (\%) &
	Download speed (kB/s) \\
	\hline
	Youku  & 39.5&23.0 & 50\\
	56.com & 5.4&6.5   & 25\\
	Tudou  & 7.3&5.9   & 50\\
	Sina.video.com & 16.3 & 1.5 & 57\\
	Tv.sohu.com & 8.8  & 1.5 & 121\\
	Ku6.com & 0.58 & 1.3  & 6.2\\
	Joy.cn  & 0.04 & 0.16 & 6.0\\
	Yinyuetai.com & 0.91  & 0.06 & 53\\
	Others & 21.1 & 60.1 & 27\\
	\hline
\end{tabu}
\end{table}

For a video flow, denote its start time by $t$, then we say that a flow is an 
\emph{OSN related video flow} if the previous flow with the same source IP came 
from an OSN in time interval $(t-\Delta,t)$. When $\Delta=1$(second), we find that 
$9.9\%$ of the video flows are OSN related video flows, which account for $25.1\%$ 
of the traffic generated by all videos. When $\Delta=2$(seconds), about $10.6\%$ of 
the video flows are OSN related video flows, which account for $26.0\%$ of the 
traffic generated by all videos. For each video website, the traffic and flow 
fractions are shown in Table \ref{tab:osn_video_traffic}. For the shortage of 
space, in the following analysis we set $\Delta=1$(second). 

\begin{table}
	\centering
	\caption{Traffic and flow fractions of OSN related videos for each online video 
	website.\label{tab:osn_video_traffic}}
	\begin{tabular}{l|rr|rr}
		\hline\hline
		\multirow{2}{*}{Website} & \multicolumn{2}{c|}{Traffic fraction (\%)} & 
		\multicolumn{2}{c}{Flow fraction (\%)} \\
		\cline{2-5}
		& $\Delta=1$ & $\Delta=2$ & $\Delta=1$ & $\Delta=2$ \\
		\hline
		Youku & 19.8 & 20.8 & 10.6 & 11.3\\
		56.com & 27.6 & 28.1 & 15.6 & 16.7\\
		Tudou & 53.9 & 55.0 & 17.1 & 18.4\\
		Sina.video.com & 21.7 & 22.9 & 32.2 & 33.4\\
		Tv.sohu.com & 23.0 & 23.2 & 19.1 & 21.0\\
		Ku6.com & 61.3 & 61.5 & 17.3 & 19.4\\
		Joy.cn & 40.9 & 41.0 & 21.1 & 21.8\\
		Yinyuetai.com & 47.8 & 47.9 & 30.4 & 31.6\\
		Others & 25.9 & 26.2 & 7.3 & 7.9\\
		\hline
	\end{tabular}
\end{table}

Next, we study the traffic properties inside OSNs. The traffic distributions of OSN 
related video flows among OSNs are shown in Table~\ref{tab:osn_traffic_stat}. We 
can see that videos flows are much more popular in Renren and Qzone, than in Sina 
and Tencent Weibo. This finding is consistent with the results shown in 
Table~\ref{tab:yk_t20}, which are obtained from the datasets collected from online 
video websites Youku and Tudou. We further show the traffic distributions of OSN 
related video flows among the video websites for each OSN website in 
Table~\ref{tab:vs_traffic}. We can see that the top-3 most popular video websites 
in each OSN are Youku, Ku6.com, and Tudou, which is also consistent with the 
results shown in Table~\ref{tab:rr_cat} that are obtained from the dataset 
collected from Renren.

\begin{table}
	\centering
	\caption{Traffic distributions of OSN related video flows among 
	OSNs.\label{tab:osn_traffic_stat}}
	\begin{tabu}{l|X[r]X[r]}
		\hline\hline
		Website & Distribution of \#bytes (\%) & Distribution of \#flows(\%)\\
		\hline
		Renren & 54.7 & 38.5 \\
		Qzone.qq.com & 24.6 & 39.9 \\
		Kaixin001.com & 0.4 & 0.9 \\
		Douban.com & 1.0 & 1.9 \\
		Sina Weibo (weibo.com) & 4.4 & 6.7 \\
		Tencent Weibo (t.qq.com) & 14.5 & 12.1 \\
		\hline
	\end{tabu}
\end{table}

\begin{table*}
	\centering
	\caption{Traffic distributions of OSN related video flows among online video  
	websites for each OSN website. \label{tab:vs_traffic}}
	\begin{tabular}{l|l|rrrrrrrrr}
		\hline\hline
		 & Website & Youku & Ku6.com & Tudou & 56.com & Sina.video.com & Joy.cn & 
		 Yinyuetai.com & Tv.sohu.com  & Others\\
		\hline
		\multirow{6}{*}{Dist. of \#bytes (\%)} & Renren & 43.9 & 1.9 & 19.2 
		& 5.4 & 9.5 & 0.1 & 0.4 & 1.4 & 18.2\\
		 & Qzone.qq.com & 17.4 & 1.1 & 15.8 & 6.7 & 8.9 & 0 & 5.0 & 13.0 & 32.0\\
		 & Kaixin001.com & 40.3 & 0 & 0 & 9.5 & 11.2 & 2.1 & 0 & 0.1 & 36.8\\
		 & Douban.com & 17.4 & 0.5 & 3.9 & 6.0 & 11.3 & 0 & 0 & 26.6 & 34.4\\
		 & Sina weibo (weibo.com) & 10.2 & 0.3 & 3.6 & 4.8 & 42.4 & 0 & 1.7 & 24.5 
		 & 12.5\\
		 & Tencent weibo (t.qq.com) & 20.6 & 0.4 & 16.7 & 11.5 & 8.5 & 0 & 1.1 & 
		 5.6 & 35.7\\
		\hline
		\multirow{6}{*}{Dist. of \#flows (\%)} & Renren & 47.7 & 3.9 & 7.1 & 
		8.2 & 3.5 & 0.5 & 0.1 & 1.2 & 27.8\\
		 & Qzone.qq.com & 9.0 & 1.3 & 13.7 & 10.7 & 3.2 & 0.3 & 0.3 & 3.5 & 58.1\\
		 & Kaixin001.com & 22.5 & 0 & 8.0 & 3.1 & 5.0 & 3.1 & 0 & 1.7 & 56.6\\
		 & Douban.com & 11.1 & 0.8 & 6.1 & 15.5 & 6.1 & 0 & 0 & 5.4 & 54.9\\
		 & Sina weibo (weibo.com) & 9.0 & 0.7 & 12.0 & 10.1 & 15.7 & 0.3 & 0.3 & 
		 5.8 & 
		 46.2\\
		 & Tencent weibo (t.qq.com) & 18.6 & 0.4 & 5.0 & 18.8 & 3.1 & 0.1 & 0 & 2.3 
		 & 51.6\\
		\hline
	\end{tabular}
\end{table*}

\section{Conclusions}\label{sec:conclusions}
We comprehensively studied video sharing between online 
video websites and online social networks from three perspective views: online 
video websites, online social networks, and network traffic of a campus network. 
The results of our study provide insights on the interplays between the two kinds 
of websites, which are summarized bellow: 

a) Many factors can affect the external sharing probability of videos in 
online video websites. The nature of a video, e.g., its category, is also an 
important factor. 

b) The popularity of a video itself in online video websites can
greatly impact on its popularity in OSNs. Videos in Renren and Qzone usually 
attract more viewers than in Sina and Tencent Weibo, which indicates the different 
natures of the two kinds of OSNs.

c) 10\% of video flows are associated with online social networks, and they account 
for 25\% of traffic generated by all videos.

%\section*{Acknowledgment}
%The research presented in this paper is supported in part by National Natural 
%Science Foundation (60574087), 
%863 High Tech Development Plan (2007AA01Z475, 2007AA01Z480, 2007AA01Z464, 
%2008AA01Z415).

\bibliographystyle{IEEEtran}
\bibliography{IEEEabrv,reference}

\end{document}